%Data da ultima versao: 22/1/97
\magnification=1200
\hsize=17 true cm
\vsize=22 true cm
\tolerance=6000
\baselineskip=21pt
\font\tiny=cmr7

\def\ref {\par \noindent \parshape=6 0cm 12.5cm
 0.5cm 12.5cm 0.5cm 12.5cm 0.5cm 12.5cm 0.5cm 12.5cm 0.5cm 12.5cm}
 
% Next 5 lines define \simless and \simgreat: "less than or approximately
% equal to" and "greater than or approximately equal to".
\newbox\grsign \setbox\grsign=\hbox{$>$} \newdimen\grdimen \grdimen=\ht\grsign
\newbox\simlessbox \newbox\simgreatbox
\setbox\simgreatbox=\hbox{\raise.5ex\hbox{$>$}\llap
     {\lower.5ex\hbox{$\sim$}}}\ht1=\grdimen\dp1=0pt
\setbox\simlessbox=\hbox{\raise.5ex\hbox{$<$}\llap
     {\lower.5ex\hbox{$\sim$}}}\ht2=\grdimen\dp2=0pt
\def\simgreat{\mathrel{\copy\simgreatbox}}

\vtop to 2.0 true cm {}
\centerline {\bf THE FeH WING-FORD BAND IN SPECTRA
OF M STARS
\footnote{$^\star$}{\tiny Observations collected at Laborat\'orio Nacional
de Astrof\'\i sica, Pico dos Dias, Brazil}}
\vskip 1.0 true cm
\centerline { Ricardo P. Schiavon, B. Barbuy \& Patan D. Singh}
\centerline {\it ripisc@atmos.iagusp.usp.br, barbuy@orion.iagusp.usp.br,
pdsingh@orion.iagusp.usp.br}
\bigskip
\bigskip
\centerline {Universidade de S\~ao Paulo, Instituto Astron\^omico 
Geof\'\i sico,
Departamento de Astronomia}

\centerline { C.P. 9638, S\~ao Paulo 01065-970, Brazil}
\vskip 3.0 true cm

\noindent Submitted to: The Astrophysical Journal.

\noindent Send proofs to: R. P. Schiavon 

\vfill\eject

\noindent {ABSTRACT}
\bigskip
\noindent We study the FeH Wing-Ford band at $\lambda\lambda$
9850 - 10200 ${\rm \AA}$ by means of the fit of synthetic spectra
to the observations of M stars, employing recent model atmospheres.
On the basis of the spectrum synthesis, we analyze the dependence of the 
band upon atmospheric parameters. FeH lines are a very sensitive surface 
gravity indicator, being stronger in dwarfs. They are also sensitive to 
metallicity (Allard \& Hauschildt 1995). The blending with CN lines, which are
stronger in giants, does not affect the response of the Wing-Ford band 
to surface gravity at low resolution (or high velocity dispersions) because 
CN lines, which are spread all along the spectrum, are smeared out at 
convolutions of FWHM $\simgreat$ 3 ${\rm \AA}$. We conclude that the Wing-Ford 
band is a suitable dwarf/giant indicator for the study of composite stellar
populations.
\bigskip
\noindent {\bf Subject Headings:} 
Stars: atmospheres, fundamental parameters, M stars - Galaxies: stellar 
content - Physical data and processes: molecular data.

\bigskip
\noindent{1. INTRODUCTION}
\vskip 1.0 true cm
Iron Hydride (FeH) is a typical signature of the atmospheres of the
coolest stars. FeH lines have been detected in spectra of M stars
(Wing \& Ford 1969), S stars (Wing 1972) and sunspots (Carrol \&
McCormack 1972; Carrol, McCormack \& O'Connor 1976). Wing \& Ford were
the first to detect a band at $\lambda\sim$ 9900${\rm \AA}$ in
the spectra of the M dwarfs Wolf 359 and Barnard's star (Gl 699).
Nordh, Lindgren \& Wing (1977) suggested that the
Wing-Ford band was due to FeH, based on the similarity between stellar
and laboratory low-resolution spectra.
This near infrared (NIR) band was unambiguously assigned to FeH by Wing,
Cohen \& Brault (1977), through the comparison of sunspot with
laboratory high-resolution spectra. As in the case of other hydrides 
(e.g. MgH, CaH), FeH lines are stronger in dwarfs than in giants.
 Whitford (1977) and Carter, Visvanathan \&
Pickles (1986) attempted to observe the Wing-Ford Band (WFB) in
integrated spectra of galaxies, in order to estimate the contribution of M
dwarfs to their integrated light, but there was no clear detection of 
the band.
More recently, Hardy \& Couture (1988), Davidge
(1991) and Couture \& Hardy (1993) detected the WFB in
integrated spectra of elliptical and lenticular galaxies.

On the laboratory side, the recent years witnessed a large improvement
in the knowledge of the structure of FeH. The first laboratory spectrum
of FeH was produced by Kleman \& $\rm \AA$kerlind (1957, unpublished),
but the rotational structure of the line spectrum was not analyzed.
Balfour, Lindgren \& O'Connor (1983) showed that the WFB
was associated to a $^4\Delta-^4\Delta$ transition. Phillips et
al. (1987) carried out a rotational analysis of the WFB,
identifying seven vibrational bands and determining rotational and
vibrational constants and spin splittings. 
Langhoff \& Bauschlicher (1988) found
theoretical evidence in favour of a $^4\Delta$ ground state
for FeH. This has been confirmed by recent laboratory results (Carter,
Steimle \& Brown 1993), which showed that the ground state of FeH is the
lower level of the transition associated to the WFB.
There is no laboratory
determination of the electronic oscillator strenght ($f_{el}$) of the
A$^4\Delta$--X$^4\Delta$ transition.

In this work we compute synthetic spectra for the range of atmospheric
parameters of M stars, employing updated model stellar photospheres
(Plez et al. 1992; Allard \& Hauschildt 1995).
These computations are compared to high resolution spectra of M
dwarfs and giants in the spectral region of the WFB. Our main
interest is to study the behavior of the WFB as a function of stellar
parameters, in order to test its usefulness as a 
discriminator between M giants and
dwarfs. In a previous paper (Schiavon
et al. 1996), we carried out a similar study applied to the NaI 
NIR doublet. 
\par In Section 2 the observations are reported.
The molecular constants
employed are given in Section 3. 
The computation of synthetic spectra is described 
in Section 4. In Section 5 the results are discussed.
 A summary is given in Section 6.
\vskip 1.0 true cm
\noindent{2. OBSERVATIONS}
\vskip 0.5 true cm
The observations were collected at the coud\'e focus of the
Boller \& Chivens 1.6m 
telescope of the  Laborat\'orio Nacional de Astrof\'\i
sica (LNA), Pico dos Dias, Brazil, in 1995 January
and August.  
 A grid of 600 l/mm and
a EEV CCD of 770$\times$1152 pixels were used, resulting in a spectral
coverage from 9850${\rm \AA}$ to 10300${\rm \AA}$, with a resolution of
0.8${\rm \AA}$.

The program stars are listed in Table 1, together with apparent visual
magnitudes, spectral types (from Gliese \& Jahreiss, 1991, for the
dwarfs, and from Hoffleit \& Warren, 1991, for the giants), 
and their effective temperatures and surface gravities, as estimated in
Section 2.1.

The spectra were reduced in the usual way; bias subtraction, 
flatfield
correction, spectrum extraction and wavelength calibration were
carried out with IRAF routines.
 Telluric absorption lines were removed through
division by the spectra of early-type stars of
high v$\sin i$. In
Figs. 1a, b we display the spectra of program dwarfs
and giants, respectively. The WFB is visible in spectra of dwarfs at
T$_{\rm eff}\sim 3700$K, but it becomes strong
only for T$_{\rm eff}\leq3500$K. In spectra of giants FeH lines are not
so clearly distinguished because of the blending with CN lines, which
are stronger in the spectra of lower gravity stars.
In spectra of 
giants cooler than M5, the TiO bands of the $\epsilon$ (E$^3\Pi$--X$^3\Delta$),
$\delta$ (b$^1\Pi$--a$^1\Delta$) and
$\phi$ (b$^1\Pi$--d$^1\Sigma$) systems (Jorgensen 1994)
become apparent and dominate the spectrum of the coolest
giant of our sample, HR 1492 (M8). 
\vskip 1.0 true cm
\noindent{\it 2.1 Atmospheric Parameters for the Dwarfs}
\vskip 0.5 true cm
Effective temperatures for the dwarfs were
estimated from the (V--K$_{\rm CIT}$) colors compiled by Leggett (1992,
hereafter L92),
adopting the relation of Jones et al. (1996), where K$_{\rm CIT}$
is defined by
Frogel et al. (1978). For two stars (Gl 357 and 842), no 
K$_{\rm CIT}$ magnitudes are available. In these cases, effective 
temperatures were
estimated from the Cousins (R--I) colors (Bessell 1989), adopting the
calibration from Bessell (1990). 

The surface gravities were estimated as
follows: from the parallaxes and V magnitudes listed by L92
and from the mass {\it vs.} M$_{\rm V}$ relation of Henry \& McCarthy
(1993) we derived stellar masses. Using K$_{\rm CIT}$ and parallaxes from
L92, we estimated M$_{\rm bol}$ from the M$_{\rm bol}$ {\it
vs.} M$_{\rm K}$ relation given by Jones et al. (1996). For Gl 357 and
842, the surface gravities were derived taking M$_{\rm V}$ from L92,
a bolometric correction estimated from Bessell (1990) and the
mass {\it vs.} M$_{\rm V}$ relation of Henry \& McCarthy (1993).

Metallicities are estimated from the stellar populations 
classification of L92 for M dwarf stars, shown in the last
column of Table 1. This classification is based on the
kinematic and photometric characteristics (in NIR colors) of a sample of
nearby red dwarfs. The stars are divided into five groups: halo stars,
 old disk/halo stars, old disk stars, old
disk/young disk stars and young disk stars.
\vskip 1.0 true cm
\noindent{\it 2.2 Atmospheric Parameters for the Giants}
\vskip 0.5 true cm
For the
giants we derived effective temperatures from the T$_{\rm eff}$ {\it
vs.} spectral type relation from Fluks et al. (1994). As there is no
reliable determination of the masses and distances to the giant stars of
Table 1, we do not estimate their surface gravities, adopting $\log g$=1.5
to all stars.
\vskip 1.0 true cm
\noindent{3. DETERMINATION OF THE MOLECULAR CONSTANTS}
\vskip 0.5 true cm
The WFB is due to the A$^4\Delta$--X$^4\Delta$ transition of
FeH. The wavelengths of FeH rotational lines for the vibrational
transitions (v',v'')=(0,0), (0,1), (0,2), (1,0), (1,1), (1,2), (2,0), (2,1) 
and (2,2) of the A$^4\Delta$--X$^4\Delta$ electronic transition were
kindly provided by Dr. J.G. Phillips (Phillips et al. 1987) in digitized 
format. 
\vskip 1.0 true cm
\noindent{\it 3.1 H\"onl-London and Franck-Condon Factors}
\vskip 0.5 true cm
The H\"onl-London
factors were computed using the code by Whiting \&
Nicholls (1973), which was kindly made available to us by Dr. J. Brown.

The Franck-Condon factors ($q_{v'v''}$) were computed with the program
for transition probabilities of molecular transitions written by Jarmain
\& McCallum (1970). The observed energy levels ($E_v$) and rotational
constants for the two participating electronic states of the FeH molecule
are from the laboratory analyses of Carter et al. (1993) and Phillips et al.
(1987). The minimum and maximum integration limits for the wavefunctions of
both electronic states were fixed at 1.42 and 6.62{\it bohr}, respectively.
Since $r'_e \sim r''_e$, strong bands lie on the $\Delta v$=0 sequence. 
For the dissociation energy, we adopted a recent laboratory determination
from Schulz \& Armentrout (1991), D$_0$=1.63eV. In
Table 2 we present  the $q_{v'v''}$ and r-centroids
for the FeH A$^4\Delta$--X$^4\Delta$ system. Since $$\sum_{v'} q_{v'0} \sim 1,$$
direct photodissociation of the molecule is not possible through the 
A$^4\Delta$ state.
\vskip 1.0 true cm
\noindent{\it 3.2 Electronic Oscillator Strength}
\vskip 0.5 true cm
We adopted an empirical value for the electronic oscillator
strength of the A$^4\Delta$--X$^4\Delta$ transition, requiring consistency 
between the model photospheres employed and the observations. 
One of our program stars (Gl 699) was studied by
Jones et al. (1996), who determined the stellar
atmospheric parameters of a number of M dwarfs from spectrum synthesis
of low resolution spectra in the NIR, curves of growth of atomic lines
and kinematic data. Their analysis was based on an improved version of 
the Allard \& Hauschildt (1995)
model photospheres. For Gl
699, they found T$_{\rm eff}$=3200K, $\log g$=5.0 and --1.5$<$[Fe/H]$<$--1.0.
We fitted the observed intensity of the WFB in Gl 699 with a synthetic
spectrum computed
 with T$_{\rm eff}$=3200K, $\log g$=5.0 and [Fe/H]=--1.0, having obtained
$f_{el}\sim 1\times 10^{-3}$. This value is one order of magnitude lower than
the value proposed by Langhoff \& Bauschlicher (1990), on the basis of
{\it ab initio} calculations.
\vskip 1.0 true cm
\noindent{\it 3.3 Dissociative Equilibrium Constant}
\vskip 0.5 true cm
The dissociative equilibrium was computed following Tsuji (1973).
The dissociative equilibrium constant of FeH as a function of 
reciprocal temperature ($\theta$=5040K/T) 
K$_{\rm FeH}$($\theta$), is not available in the work of Tsuji (1973). 
We computed K$_{\rm FeH}$(T) from equation (7) of Tatum (1966). 
The molecular partition
function as a function of temperature, Q$_{\rm FeH}$(T), was computed from
equation (15) of Tatum (1966), taking the electronic terms, the
rotational, vibrational and anharmonicity constants from Phillips et al.
(1987). For the 
partition function of Fe as a function
of temperature, we adopted the polynomial fit given by Irwin (1981).
The function K$_{\rm FeH}$($\theta$) thus obtained was fitted by a
fourth order polynomial as follows:
$$
\log {\rm K_{FeH}(\theta) = \sum_{i=0}^4 a_i\theta^i}
$$
where (${\rm a_0,a_1,a_2,a_3,a_4}$)=(12.867,--3.8754,1.0326,--0.22476,0.01775).
Our calculations show that the region of the WFB in the spectra of M
stars is dominated by FeH, CN and atomic lines. The CN lines due to the 
(1,0), (2,1) and (3,2) vibrational
bands of the A$^2\Pi$ $-$ X$^2\Sigma$ transition were taken from the
work of Davis \& Phillips (1963).
\vskip 1.0 true cm
\noindent{4. SPECTRUM SYNTHESIS}
\vskip 0.5 true cm
The synthetic spectra were computed in the interval 
$\lambda\lambda 9850 - 10200{\rm \AA}$. The spectrum synthesis code used
is described in Barbuy (1982). The model photospheres adopted are from Allard 
\& Hauschildt (1995) for the dwarfs and from Plez et al. (1992) for the giants.
The atomic line list was taken from Swensson et al. (1973), with oscillator 
strengths and damping constants obtained by fitting the solar spectrum 
(Kurucz et al. 1984), adopting the solar model photosphere of Kurucz (1992).
For the line profiles we adopted the Hjerting function.
\vskip 1.0 true cm
\noindent{\it 4.1 Dwarfs}
\vskip 0.5 true cm
We computed synthetic spectra for a grid of 157 model photospheres
in the range of atmospheric parameters 
2700${\rm \leq T_{eff}\leq}$4000K, $4.0\leq\log g\leq 5.5$ and $-1.5{\rm 
\leq[Fe/H]\leq}+0.5$. 
In Fig. 2 are shown synthetic spectra computed for 
T$_{\rm eff}$=3200K, $\log g$=5.0 and solar 
metallicity, with:
 a) CN lines only,
b) atomic lines only and c) FeH, CN and atomic lines. 

Adopting the atmospheric parameters given in Table 1, we obtained a good
match between synthetic and observed spectra for 8 of the program dwarf
stars. Two of the remaining 5
stars (Gl 357 and 842) are not included in the list of L92, 
so that we do not have estimates for their metallicities. For
the other three stars (Gl 1, 190 and 285), the synthetic spectra did not 
match the observed ones. 

In Fig. 3a we show, as a typical example, the comparison of spectrum synthesis
and observation for Gl 699. 
The good agreement between synthetic and observed spectra for the
majority of the program stars validates our choice of the
empirical value of the electronic strength for the A$^4\Delta$--
X$^4\Delta$  transition of FeH.

For Gl 1, 190 and 285, we improved the estimate of the atmospheric
parameters in three steps:

{\it i)} For each star we computed the {\it rms} deviation of each synthetic 
spectrum of our grid relative to the observed spectrum 
and selected the best 20\% synthetic spectra.

{\it ii)} We measured the depth ratios of a set of lines selected on
the basis of their sensitivity to atmospheric parameters. The lines used 
are listed in Table 3, together
with the identification of the contributors to the absorption. 
 All the line 
ratios present a strong sensitivity to at least one of the three atmospheric
parameters (Figs. 4a,b), so that the set of 10 pairs can be used as
a further constraint to select the atmospheric parameters. 
 We require that the difference between the observed and
synthetic line ratios be of the order of the uncertainties in the measurements 
(usually $\sim$ 10-20\%). The uncertainties are estimated using the expression 
of Gray \& Johansson (1991).

{\it iii)} The resulting values of T$_{\rm eff}$, $\log g$ and [Fe/H] 
are finally
compared to those of Table 1, requiring that $\Delta$T$_{\rm eff}$ $<$
$\pm$300K, $\Delta$log g $<$ $\pm$0.5
 and $\Delta$[Fe/H] $<$ $\pm$0.5. 

After the selection process, there remained five sets of distinct
atmospheric parameters per star, on average. 
These final atmospheric parameters were averaged 
and the resulting
values are shown in Table 4. In order to check the
reliability of this method, it was applied to the 8 stars for
which the synthetic spectra based on the atmospheric parameters
of Table 1 presented a good fit to the observations. By
comparing the values of Tables 1 and 4, it is seen that the
differences in T$_{\rm eff}$, $\log g$ and [Fe/H] for these 8
stars are smaller than 150K, 0.2 and 0.4 dex, respectively. In Figs.
3b,c we show the comparison of synthetic and observed spectra for Gl 1
and 285, which are, respectively, the hotter and cooler star in our
sample. 
\vskip 1.0 true cm
\noindent{\it 4.2 Giants}
\vskip 0.5 true cm
In Fig. 5 we show the spectrum synthesis for the giant
star HR 832 (M4), where the atmospheric
parameters adopted are (T$_{\rm eff}$,$\log g$,[Fe/H]) =
(3400K,1.5,0.0). 

CN lines become important in the spectra of giants, as 
illustrated in Fig. 6, where the separate contributions of molecular and atomic
lines are shown for stellar parameters
T$_{\rm eff}$=3600K, $\log g$=1.5 and solar metallicity (compare with
Fig. 2, for the case of dwarfs). Because CN lines are more important
in the spectra of giants, they change the response of the equivalent
width of the WFB to atmospheric parameters.
\vskip 1.0 true cm
\noindent{5. THE WING-FORD BAND AS A FUNCTION OF ATMOSPHERIC PARAMETERS
AND SPECTRAL RESOLUTION}
\vskip 0.5 true cm
The equivalent width of the WFB was measured in the 157 synthetic spectra
of the grid,
in the interval $\lambda\lambda$ 9896-9980 ${\rm\AA}$, with continuum
points located at the endpoints of this interval. 

The equivalent width
 measured in spectra of giants has an important contribution from
CN lines (Fig. 6). 
This contribution is a strong function of the resolution of the
spectrum. 
In Fig. 7 we show the synthetic spectrum due to CN only, for a giant 
of T$_{\rm eff}$=3600K convolved with gaussians of FWHM=0.8 and 5${\rm \AA}$.
From this Fig. it is seen that, since CN lines are spread all
along the spectral region, showing no pronounced features, their
presence is diluted at lower resolutions. Therefore, the main effect of
CN lines at low resolution is the lowering of the pseudo-continuum adjacent
to the WFB.
Therefore, we convolved our grid of synthetic spectra with gaussians of
different FWHM, in order to study the effect of resolution upon the
equivalent width measurements.
\smallskip
\par {\it 5.1 Behavior of WFB at High Resolution}
\smallskip
\par In Fig. 8 we show the equivalent width of the WFB  as a function of
effective temperature and surface gravity, computed for FWHM=0.8${\rm \AA}$. 
As expected, the WFB is very
sensitive to both parameters. There is a clear distinction between the
behavior of giants and dwarfs. In dwarfs FeH lines are predominant whereas
CN lines are stronger in giants.

In the dwarf regime, the WFB behaves as expected, for T$_{\rm eff}$
$\leq$ 3300K, it becomes stronger for higher $\log g$. (It has to be noted 
that for higher temperatures, there is an inversion of this trend for the 
models at
$\log g$=4.0 and 4.5. The same inversion happens for the models at $\log
g$=4.5 and 5.0 at T$_{\rm eff}$$\sim$ 3500K. This inversion is presumably due 
to the effect pointed out by Brett (1995, his Fig. 3), concerning the 
decreasing
importance of convective transport for decreasing gravities, and the
influence of convection on the temperature gradient of the photosphere.
The effect does not appear in lower T$_{\rm eff}$ models because the
efficiency of convective transport reaches a maximum at T$_{\rm
eff}\sim$ 3200K).

In Fig. 9 we show the behavior of the WFB equivalent width as a
function of effective temperature and metallicity for dwarfs.
(This plot shows the unexpected behavior of
stronger FeH lines at lower metallicities. This
effect was already pointed out by Allard \& Hauschildt (1995) and is
related to the increased gas pressure of metal poor photospheres, at a
given optical depth, which overcompensates the lower particle number
density; the same
result is obtained using Kurucz (1992) model photospheres for T$_{\rm
eff}$=3500K. For T$_{\rm eff}>$3300K, the trend begins to change for
higher metallicities, so that the band assumes the usual behavior for
T$_{\rm eff}\sim$4000K).

In the giants CN lines dominate the WFB equivalent width; 
CN lines are stronger for lower $\log g$ and higher T$_{\rm eff}$ 
(the latter being valid for T$_{\rm eff}$ $<$ 4400 K, cf. Milone \& Barbuy 
1994).
\smallskip
\par {\it 5.2 Behavior of WFB at Low Resolution}
\smallskip
In Figs. 10a to 10d we show the dependence
of the WFB on T$_{\rm eff}$ and $\log g$ for spectra convolved with gaussians
of increasing FWHM. For FHWM=2${\rm \AA}$,
the dependence of WFB on gravity is not significantly different from what is
seen in Fig. 8. As convolution increases, the
pseudo-continuum level decreases in the spectra of giants, 
leading to lower WFB equivalent widths. The effect is more important for
lower resolution (higher FWHM) and stronger CN lines. Hence, while
the WFB equivalent width in dwarfs is almost invariant with convolution, 
it is sensibly lower in giants for higher FHWM. Therefore, the WFB is a
suitable dwarf/giant discriminator at lower resolutions, where the blend of CN
inside the interval $\lambda\lambda$ 9896-9980${\rm \AA}$
is compensated by the lowering of the adjacent 
pseudo-continuum due to unresolved CN lines.

Stellar velocity dispersions in elliptical galaxies are around   
50 $<$ $\sigma$ $<$ 350 km/s (e.g. Ore et al. 1991), which corresponds
to 0.9 $<$ FWHM $<$ 6 ${\rm \AA}$. 
Based on this discussion, 
it appears that the WFB is an appropriate indicator of cool dwarf populations 
for galaxies with $\sigma\simgreat200$km/s. For galaxies of lower $\sigma$
the WFB observed in lower resolution can also be used as a dwarf-giant
discriminator.
\vskip 1.0 true cm
\noindent{6. SUMMARY}
\vskip 0.5 true cm
We computed synthetic spectra for a range of atmospheric parameters, using
 state-of-the-art model photospheres, in the interval 
$\lambda\lambda 9850-10200{\rm \AA}$. This region includes the Wing-Ford band 
(WFB) of FeH, which is a well known surface gravity indicator.

Based on a set of atmospheric parameters derived from photometric indices and
kinematic population classifications, we compared our synthetic spectra 
with observations of M dwarf stars,
obtaining a satisfactory match. 

\par The equivalent width of the WFB was studied as a function of effective
temperature, surface gravity, metallicity and spectral resolution.
FeH lines are dominant in the spectra of
dwarfs.
For dwarfs cooler than 3500K, the dependence of the equivalent width of
the WFB on $\log g$ follows the expected trend, being higher for higher
$\log g$ and lower T$_{\rm eff}$. 
\par CN lines become dominant in spectra of giants, however
at convolutions of FWHM $\simgreat$ 3 ${\rm \AA}$ the CN lines, which
are spread all along the spectrum are smoothed out, with the main
effect of lowering the  pseudo-continuum.
The same effect occurs for velocity dispersions of stars in galaxies
where
$\sigma$ $\simgreat$ 200km/s, i.e., the CN lines present
in giants are smeared out.

\par We conclude that the Wing-Ford is
a suitable dwarf/giant indicator for the study of stellar populations
in external galaxies.
\bigskip

\bigskip
\noindent The authors are indebted to B. Plez and F. Allard for kindly
providing their model atmospheres, Dr. J. Phillips for the line list of FeH
and Dr. J. Brown for the code for the computation of H\"onl-London factors. We
also thank the referee, P. Hauschildt, for his suggestions on the first version
of the manuscript. The calculations were carried out in a DEC Alpha 3000/700 
workstation provided by Fapesp. RPS acknowledges Fapesp PhD fellowship No. 
93/2177-0. Partial financial support from CNPq is also acknowledged.
\vfill\eject
\noindent{REFERENCES}
\vskip 1.0 true cm
\ref Allard, F. \& Hauschildt, P.H., 1995, ApJ, 445, 433.
\ref Balfour, W.J., Lindgren, B. \& O'Connor, S., 1983, Chem.Phys.Lett., 
96, 251.
\ref Barbuy, B., 1982, PhD thesis, Universit\'e de Paris VII.
\ref Bessell, M.S., 1990, A\&AS, 83, 357.
\ref Bessell, M.S., 1991, AJ, 101, 662.
\ref Brett, J.M., 1995, A\&A, 295, 736.
\ref Carrol, P.K. \& McCormack, P., 1972, ApJ, 177, L33.
\ref Carrol, P.K., McCormack, P. \& O'Connor S., 1976, ApJ, 208, 903.
\ref Carter, R.T., Steimle, T.C. \& Brown, J.M., 1993, J.Chem.Phys, 99, 
3166.
\ref Carter, D., Visvanathan, N. \& Pickles, A.J., 1986, ApJ, 311, 637.
\ref Couture, J. \& Hardy, E., 1993, ApJ, 406, 142.
\ref Davis, S.P. \& Phillips, J.G., 1963, The red system (A$^2\Pi$
$-$ X$^2\Sigma$) of the CN molecule, Univ. California Press.
\ref Fluks, M.A., Plez, B., Th\'e, P.S., de Winter, D., Westerlund, B.E. \&
Steenman, H.C., 1994, A\&AS, 105, 311.
\ref Frogel, J.A., Persson, S.E., Aaronson, M. \& Matthews, K.
1978, ApJ, 220, 75.
\ref Gliese, W \& Jahreiss, H., 1991, Preliminary Version of the Third
Catalogue of Nearby Stars, Astronomisches Rechen-Institute Heidelberg, Germany. 
\ref Gray, D.F., Johanson, H.L., 1991, PASP, 103, 439.
\ref Hardy, E. \& Couture, J., 1988, ApJ, 325, L29.
\ref Henry, T.J. \& McCarthy, Jr., D.W., 1993, AJ, 106, 773.
\ref Hoffleit, D. \& Warren Jr., W.H., 1991, Preliminary Version of the
Bright Star Catalogue, 5th Revised Edition.
\ref Irwin, A.W., 1981, ApJS, 45, 621.
\ref Jarmain, W.R. \& McCallum, J.C., 1970, TRAPRB, University of Western
Ontario, Dept. of Physics, Ontario, Canada.
\ref Jones, H.R.A., Longmore, A.J., Allard, F. \& Hauschildt, P.H., 1996, 
MNRAS, 280, 77.
%\ref Jorgensen, U.G., 1993, in Molecules in the Stellar Environment, IAU
%Colloquium No. 146, ed. U.G. Jorgensen, Springer-Verlag.
\ref Jorgensen, U.G.: 1994, A\&A, 284, 179.
\ref Kurucz, R., 1992, in The Stellar Populations of Galaxies,
IAU Symp. 149, eds. B. Barbuy \& A. Renzini, Kluwer Acad. Press, 225
\ref Kurucz, R.L., Furenlid, I., Brault, J., Testerman, L., 1984,
Solar Flux Atlas from 296 to 1300 nm (Sunspot, N.M., National Solar
Observatory).
\ref Langhoff, S.R. \& Bauschlicher, C.W.Jr., 1988, J.Chem.Phys., 89, 
2160.
\ref Langhoff, S.R. \& Bauschlicher, C.W.Jr., 1990, J.Mol.Spectrosc., 141, 243.
\ref Leggett, S.K., 1992, ApJS, 82, 351 (L92).
\ref Milone, A. \& Barbuy, B., 1994, A\&AS, 108, 449.
\ref Nordh, H.L., Lindgren, B. \& Wing, R.F., 1977, A\&A, 56, 1.
\ref Ore, C.D., Faber, S.M., Jesus, J., Stoughton, R., 1991, ApJ, 366, 38
\ref Plez, B., Brett, J.M. \& Nordlund, A., 1992, A\&A, 256, 551.
\ref Phillips, J.G., Davis, S.P., Lindgren, B. \& Balfour, W.J., 1987, 
ApJS, 65, 721.
\ref Schiavon, R.P., Barbuy, B., Rossi, S.C.F. \& Milone, A., 1996, ApJ, in
press.
\ref Schultz, R.H. \& Armentrout, P.B., 1991, J.Chem.Phys., 94, 2262.
\ref Spinrad, H. \& Taylor, B.J., 1971, ApJS, 22, 445.
\ref Swensson, J.W., Benedict, W.S., Delbouille, L. \& Roland, G.,
1973,  M\'em. Soc. R. Sci. Li\`ege Special, vol. 5.
\ref Tatum, J.B., 1966, Pub. Dom. Ap. Obs. Victoria, 13, 1.
\ref Tsuji, T., 1973, A\&A, 23, 411.
\ref Whitford, A.E., 1977, ApJ, 21, 527.
\ref Whiting, E.E. \& Nicholls, R.W., 1973, ApJS, 27, 1.
\ref Wing, R.F., 1972, Mem. Soc. R. Sci. Li\`ege, Ser.3, 123.
\ref Wing, R.F., Cohen, J. \& Brault, J.W., 1977, ApJ, 216, 659.
\ref Wing, R.F. \& Ford, W.K., 1969, PASP, 81, 527.
\vfill\eject

\baselineskip=15pt
\noindent {\bf Table 1 - Program Stars.}
Visual Magnitudes, spectral types, effective temperatures, surface gravities
and population class. The population classification was taken from Leggett 
(1992): halo stars
(H), old disk/halo stars (O/H), old disk stars (OD), old
disk/young disk stars (O/Y)
and young disk stars (YD). Metallicities for each class are as follows:
[Fe/H]$_{\rm H}\sim$--1.0, [Fe/H]$_{\rm O/H}\sim$--1.0, 
[Fe/H]$_{\rm OD}\sim$--0.5, [Fe/H]$_{\rm O/Y}\sim$--0.5, 
[Fe/H]$_{\rm YD}\sim$0.0. For the derivation of the other atmospheric
parameters, see Section 2.1.

$$\vbox{\halign to \hsize{\tabskip 1em plus2em$
#\hfil$&
$ #\hfil$&
$ #\hfil$&
 $#\hfil$&
$ #\hfil$&
 $#\hfil$&
 $ #\hfil\quad$&
$ #\hfil $ &
 $ #\hfil$&
$# \hfil $$\cr
\noalign{\hrule\vskip 0.2cm}
{\rm star} &
\hidewidth {\rm V} \hidewidth  &
\hidewidth {\rm Sp.T.} \hidewidth &
\hidewidth {\rm T_{eff} (K)} \hidewidth &
\hidewidth {\rm log\;g} \hidewidth &
\hidewidth {\rm Class} \hidewidth & \cr
\noalign{\vskip 0.2cm}
\noalign{\hrule\vskip 0.2cm}
\noalign{\vskip 0.2cm}
&&& {\rm Dwarfs} &&& \cr
{\rm Gl\;1}& 8.54 & {\rm M4} & 3460 & 4.9 & {\rm H} \cr
{\rm Gl\;84} & 10.19 & {\rm M3} & 3320 & 4.7 & {\rm OD} \cr
{\rm Gl\;190} & 10.30 & {\rm M4} & 3180 & 4.3 & {\rm YD} \cr
{\rm Gl\;229} & 11.2 & {\rm M1e} & 3480 & 4.6 & {\rm Y/O} \cr
{\rm Gl\;273} & 9.85 & {\rm M3.5} & 3170 & 4.8 & {\rm OD} \cr
{\rm Gl\;285} & 11.2 & {\rm M4.5e} & 3080 & 4.6 & {\rm YD} \cr
{\rm Gl\;357} & 10.92 & {\rm M3} & 3410 & 4.6 &  \cr
{\rm Gl\;581} & 10.56 & {\rm M5} & 3240 & 4.9 & {\rm YD} \cr
{\rm Gl\;699} & 9.55 & {\rm M5} & 3160 & 5.1 & {\rm O/H} \cr
{\rm Gl\;752\;A} & 9.11 & {\rm M3.5e} & 3310 & 4.7 & {\rm Y/O} \cr
{\rm Gl\;832} & 8.67 & {\rm M1} & 3400 & 4.8 & {\rm O/H} \cr
{\rm Gl\;842} & 9.75 & {\rm M2} & 3640 & 4.7 &  \cr
{\rm Gl\;876} & 10.17 & {\rm M5} & 3140 & 4.7 & {\rm YD} \cr
&&& {\rm Giants} &&& \cr
{\rm HR 625} & 6.10 & {\rm M2} & 3740 & & \cr
{\rm HR 722} & 6.41 & {\rm M5} & 3430 & & \cr
{\rm HR 832} & 6.90 & {\rm M4} & 3570 & & \cr
{\rm HR 1492} & 5.40 & {\rm M8} & 2890 & & \cr
{\rm HR 1693} & 5.68 & {\rm M6} & 3310 & & \cr
{\rm HR 8128} & 5.28 & {\rm M3} & 3670 & & \cr
\noalign{\vskip 0.01cm}  \cr}
\hrule}$$
\vfill\eject

\noindent {\bf Table 2 - Franck-Condon factors (first entry) and r-centroids
(second entry) of the A$^4\Delta$--X$^4\Delta$ transition of FeH}

$$\vbox{\halign to \hsize{\tabskip 1em plus2em$
#\hfil$&
$ #\hfil$&
$ #\hfil$&
$ #\hfil$&
$ #\hfil$&
$# \hfil $$\cr
\noalign{\hrule\vskip 0.2cm}
 &
\hidewidth v'' \hidewidth &
\hidewidth 0 \hidewidth  &
\hidewidth 1 \hidewidth &
\hidewidth 2 \hidewidth \cr
\noalign{\vskip 0.2cm}
\noalign{\hrule\vskip 0.2cm}
\noalign{\vskip 0.2cm}
v' & & & & \cr
 & & & & \cr
0  & & 8.33\times 10^{-1} & 1.56\times 10^{-1} & 1.06\times 10^{-2} \cr
   & & 1.671 & 1.938 & 2.180 \cr
 & & & & \cr
1 & & 1.46\times 10^{-1} & 5.52 \times 10^{-1}  & 2.70 \times 10^{-1}  \cr
   & & 1.456 & 1.720 & 1.992 \cr
 & & & & \cr
2 & & 1.73 \times 10^{-2} & 2.16 \times 10^{-1}  & 2.99 \times 10^{-1}  \cr
  & & 1.252 & 1.511 & 1.782 \cr
\noalign{\vskip 0.01cm}  \cr}
\hrule}$$
\vfill\eject

\noindent {\bf Table 3 - Line Pairs }

$$\vbox{\halign to \hsize{\tabskip 1em plus2em$
#\hfil$&
$ #\hfil$&
$ #\hfil$&
$# \hfil $$\cr
\noalign{\hrule\vskip 0.2cm}
\lambda_{\rm obs} &
\hidewidth {\rm Identification} \hidewidth  &
\hidewidth \chi_{\rm exc (eV)} / (v',v'')Branch-J \hidewidth \cr
\noalign{\vskip 0.2cm}
\noalign{\hrule\vskip 0.2cm}
\noalign{\vskip 0.2cm}
9899.1 & {\rm FeH}\,9899.22 & (0,0)R-15 \cr
       & {\rm FeH}\,9899.33 & (0,0)P-10 \cr
       & {\rm CN}\,9899.16 & (1,0)P_1-55 \cr
9900.9 & {\rm CrI}\,9900.95 & 2.99 \cr
       & {\rm CN}\,9900.79 & (3,2)Q_1-34 \cr
&&\cr
9926.3 & {\rm FeH}\,9926.41 & (0,0)R-12\cr
9927.4 & {\rm TiI}\,9927.38 & 1.88 \cr
       & {\rm FeH}\,9927.37 & (0,0)R-6 \cr
       & {\rm FeH}\,9927.37 & (0,0)R-10 \cr
       & {\rm CN}\,9927.49 & (3,2)Q_1-36 \cr
&&\cr
9931.7 & {\rm CaII}\,9931.35 & 7.51 \cr
       & {\rm FeH}\,9931.59 &(0,0) R-7 \cr
       & {\rm FeH}\,9931.79 & (0,0)R-3 \cr
9933.3 & {\rm SI}\,9932.37 & 8.41 \cr
       & {\rm FeH}\,9933.29 & (0,0)R-13 \cr
&&\cr
9959.6 & {\rm FeI}\,9959.17 & 4.07 \cr
       & {\rm FeH}\,9959.63 & (0,0)R-7 \cr
9961.1 & {\rm NaI}\,9961.25 & 3.62 \cr
       & {\rm FeH}\,9961.14 & (0,0)P-16 \cr
&&\cr
10007.0 & {\rm FeI}\,10007.32 & 3.02 \cr
10010.3 & {\rm FeH}\,10010.45 & (0,0)R-9 \cr
&&\cr
10010.3 & {\rm FeH}\,10010.45 & (0,0)R-9 \cr
10012.1 & {\rm FeI}\,10012.20 & 5.07 \cr
\noalign{\vskip 0.01cm}  \cr}
\hrule}$$
\vfill\eject

\noindent {\bf Table 3 - Continuation}

$$\vbox{\halign to \hsize{\tabskip 1em plus2em$
#\hfil$&
$ #\hfil$&
$ #\hfil$&
$# \hfil $$\cr
\noalign{\hrule\vskip 0.2cm}
\lambda_{\rm obs} &
\hidewidth {\rm Identification} \hidewidth  &
\hidewidth \chi_{\rm exc (eV)} / Branch-J \hidewidth \cr
\noalign{\vskip 0.2cm}
\noalign{\hrule\vskip 0.2cm}
\noalign{\vskip 0.2cm}
10033.1 & {\rm FeI}\,10032.89 & 5.50 \cr
        & {\rm FeH}\,10033.07 & (0,0)R-22 \cr
10035.7 & {\rm FeH}\,10035.60 & (0,0)R-26 \cr
        & {\rm CN}\,10035.77 & (1,0)Q_1-68 \cr
&&\cr
10074.1 & {\rm FeH}\,10074.04 & (0,0)R-22 \cr
        & {\rm CN}\,10074.07 & (3,2)Q_1-45 \cr
10075.1 & {\rm FeH}\,10075.22 & (0,0)P-17 \cr
        & {\rm FeH}\,10075.22 & (0,0)P-6 \cr
&&\cr
10187.7 & {\rm FeH}\,10187.66 & (0,0)P-11 \cr
        & {\rm CN}\,10187.86 & (2,1)Q_1-63 \cr
10189.3 & {\rm TiI}\,10189.02 & 1.46 \cr
&&\cr
10194.8 & {\rm FeI}\,10195.12 & 2.73 \cr
        & {\rm FeH}\,10194.71 & (0,0)P-11 \cr
10196.3 & {\rm FeH}\,10196.22 & (0,0)P-10 \cr
        & {\rm CN}\,10196.47 & (2,1)P_2-56 \cr
\noalign{\vskip 0.01cm}  \cr}
\hrule}$$
\vfill\eject

\noindent {\bf Table 4 - Best Set of Atmospheric Parameters for Program M
Dwarfs}

$$\vbox{\halign to \hsize{\tabskip 1em plus2em$
#\hfil$&
$ #\hfil$&
$ #\hfil$&
 $#\hfil$&
$# \hfil $$\cr
\noalign{\hrule\vskip 0.2cm}
{\rm Star} &
\hidewidth {\rm T_{eff}(K)} \hidewidth  &
\hidewidth {\rm log\;g} \hidewidth &
\hidewidth {\rm [Fe/H] } \hidewidth \cr
\noalign{\vskip 0.2cm}
\noalign{\hrule\vskip 0.2cm}
\noalign{\vskip 0.2cm}
{\rm Gl\;1}& 3650 & 4.9 & -1.1 \cr
{\rm Gl\;84} & 3210 & 4.9 & -0.6 \cr
{\rm Gl\;190} & 2900 & 4.4 & -0.2 \cr
{\rm Gl\;229} & 3330 & 4.7 & -0.2 \cr
{\rm Gl\;273} & 3180 & 4.8 & -0.8 \cr
{\rm Gl\;285} & 2830 & 5.0 & -0.3 \cr
{\rm Gl\;357} & 3190 & 4.7 & -1.0  \cr
{\rm Gl\;581} & 3230 & 5.1 & -0.1 \cr
{\rm Gl\;699} & 3260 & 5.2 & -0.9 \cr
{\rm Gl\;752\;A} & 3370 & 4.8 & -0.8 \cr
{\rm Gl\;832}  & 3320 & 4.7 & -0.7 \cr
{\rm Gl\;842} & 3410 & 4.8 & -0.6 \cr
{\rm Gl\;876} & 3130 & 4.9 & -0.4 \cr
\noalign{\vskip 0.01cm}  \cr}
\hrule}$$

\vfill\eject
\baselineskip=21pt

\centerline{\bf FIGURE CAPTIONS}

\noindent{\bf Figure 1a:} Spectra of program M dwarfs showing the Wing-Ford band
in high resolution ($\sim\lambda\lambda 9900 - 10000{\rm \AA}$). Effective
temperatures range between $\sim$ 2800K (Gl 285) to $\sim$ 3700K (Gl 1).

\noindent{\bf Figure 1b:} Spectra of program M giants. Effective
temperatures range between $\sim$ 2900K (HR 1492) to $\sim$ 3700K (HR 625).

\noindent{\bf Figure 2:} Separate contribution of CN, atomic and FeH lines for
the spectrum of a dwarf M star.

\noindent{\bf Figure 3a:} Spectrum synthesis of Gl 699 (Barnard's Star).

\noindent{\bf Figure 3b:} Spectrum synthesis of Gl 1.

\noindent{\bf Figure 3c:} Spectrum synthesis of Gl 285.

\noindent{\bf Figure 4a:} Line depth ratio of a line pair of Table 3 as a
function of T$_{\rm eff}$ and $\log g$, computed for solar metallicity.

\noindent{\bf Figure 4b:} Line depth ratio of a line pair of Table 3 as a
function of T$_{\rm eff}$ and metallicity, computed for $\log g$=5.0.

\noindent{\bf Figure 5:} Spectrum synthesis of HR 832.

\noindent{\bf Figure 6:} Separate contribution of CN, atomic and FeH lines for
the spectrum of a giant M star.

\noindent{\bf Figure 7:} CN line spectrum for a giant of T$_{\rm eff}$=3600K,
$\log g$=1.0 and solar metallicity, convolved with gaussians of FWHM=0.8${\rm
\AA}$ (dotted line) and 5.0${\rm \AA}$ (solid line).

\noindent{\bf Figure 8:} Equivalent width of the Wing-Ford band as a function
of T$_{\rm eff}$ for different gravities. The synthetic spectra were convolved
with a gaussian of FWHM=0.8${\rm \AA}$.

\noindent{\bf Figure 9:} Equivalent width of the Wing-Ford band as a function
of T$_{\rm eff}$ for different metallicities. The synthetic spectra were
convolved with a gaussian of FWHM=0.8${\rm \AA}$.

\noindent{\bf Figure 10:} Equivalent width of the Wing-Ford band as a function
of T$_{\rm eff}$ for different gravities. Each panel corresponds to a distinct
convolution of the synthetic spectra: a) FWHM=2.0${\rm \AA}$, b) 3.0${\rm
\AA}$, c) 4.0${\rm \AA}$ and d) 5.0${\rm \AA}$. The discrimination between
dwarfs and giants increases for increasing FWHM.

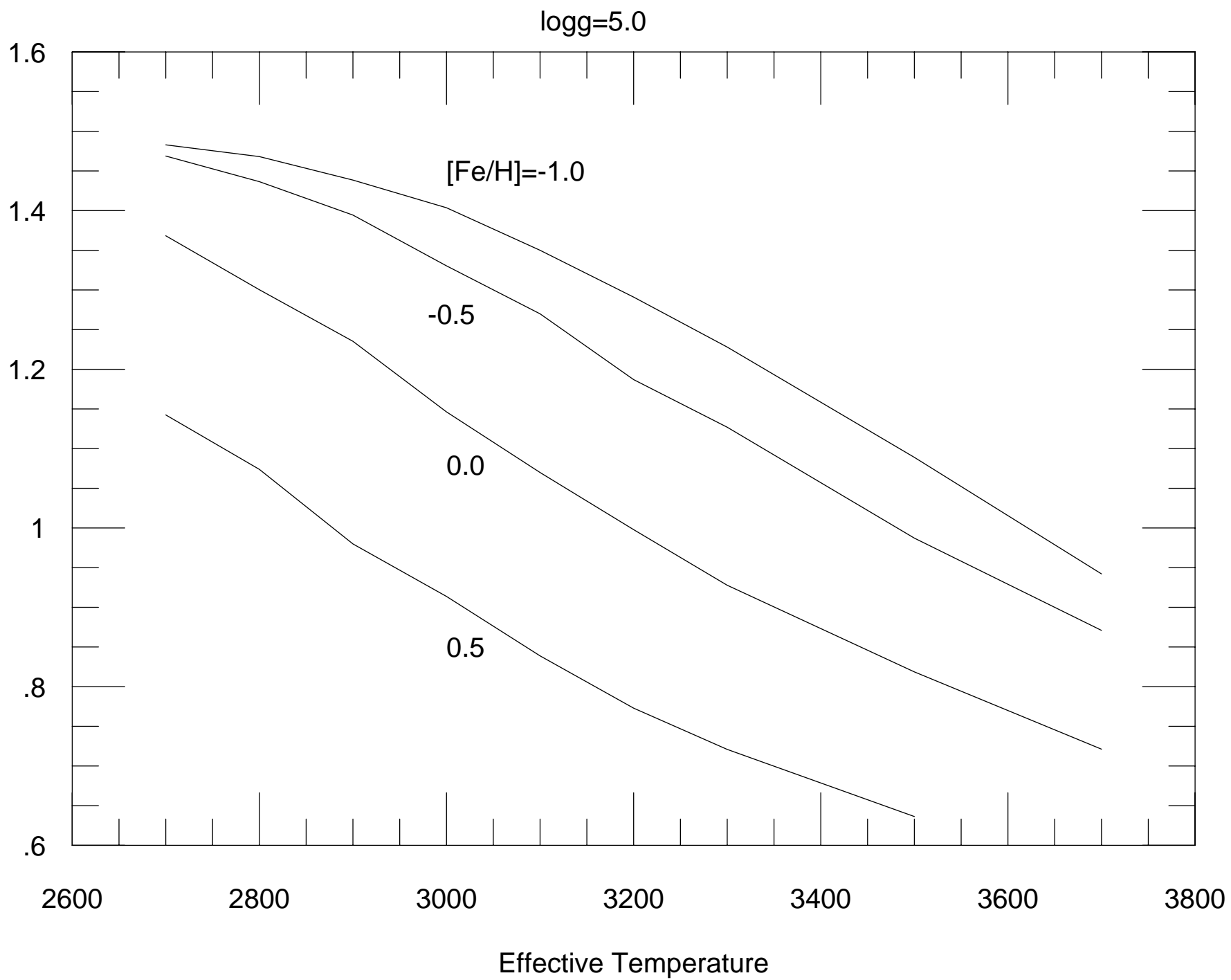
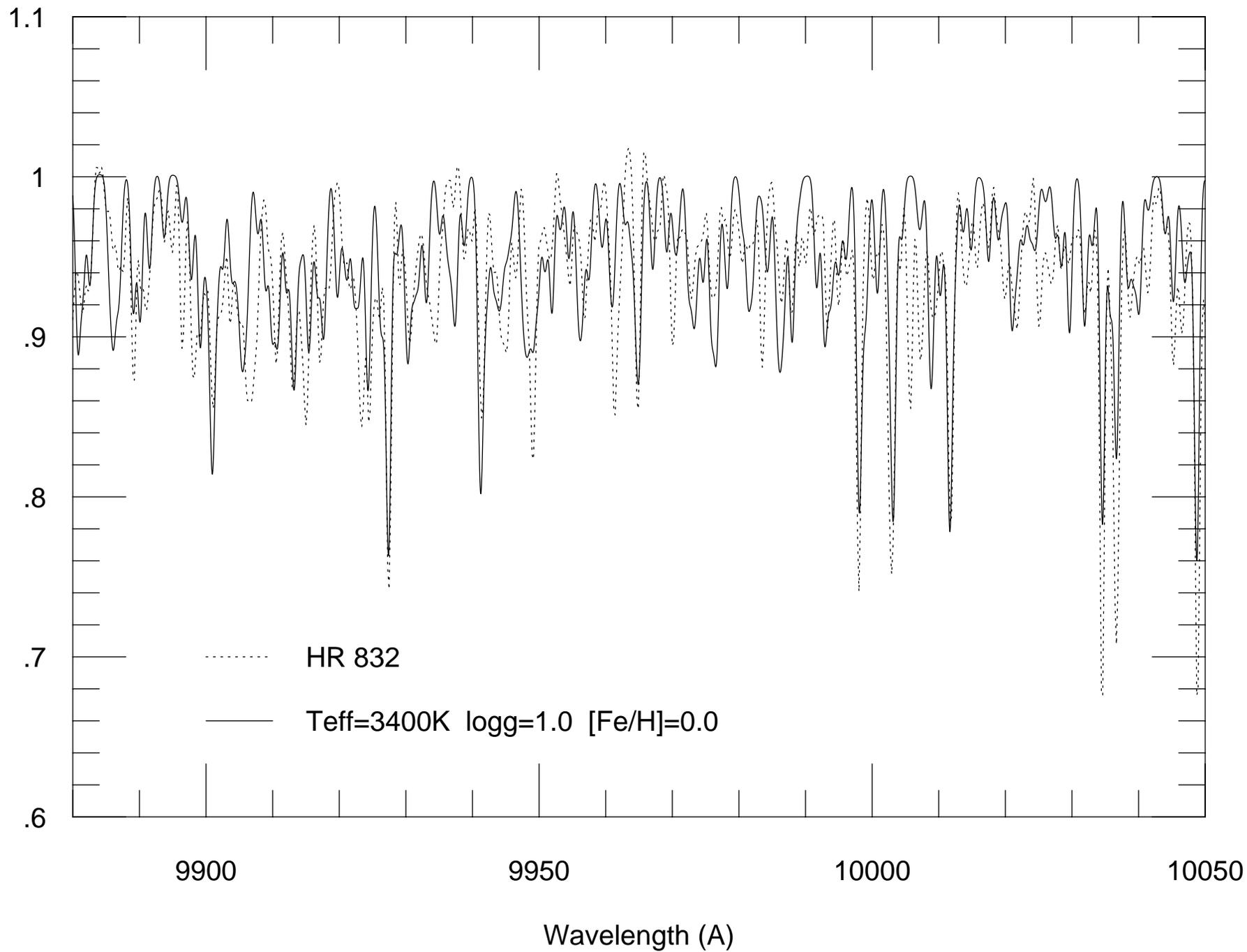
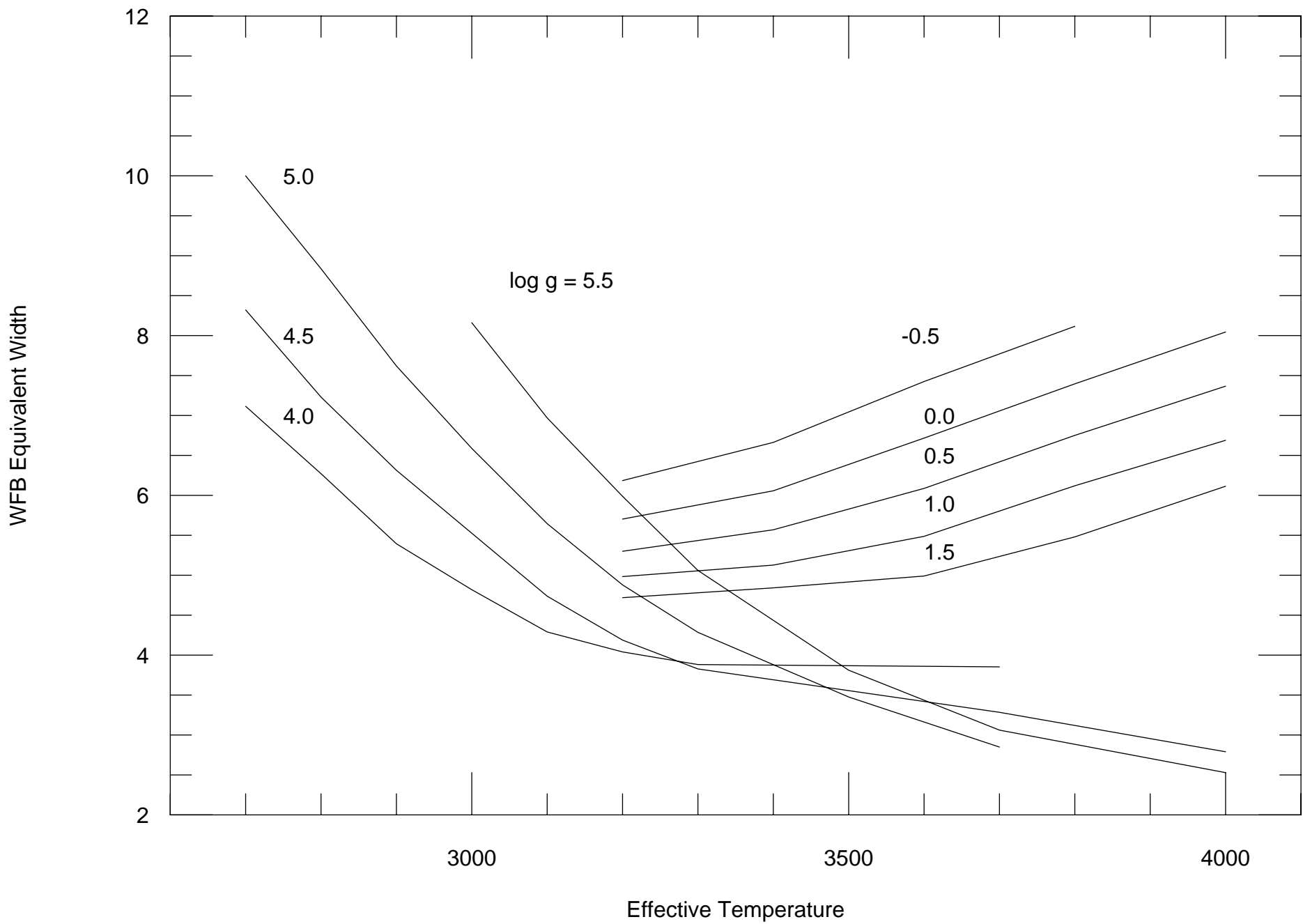
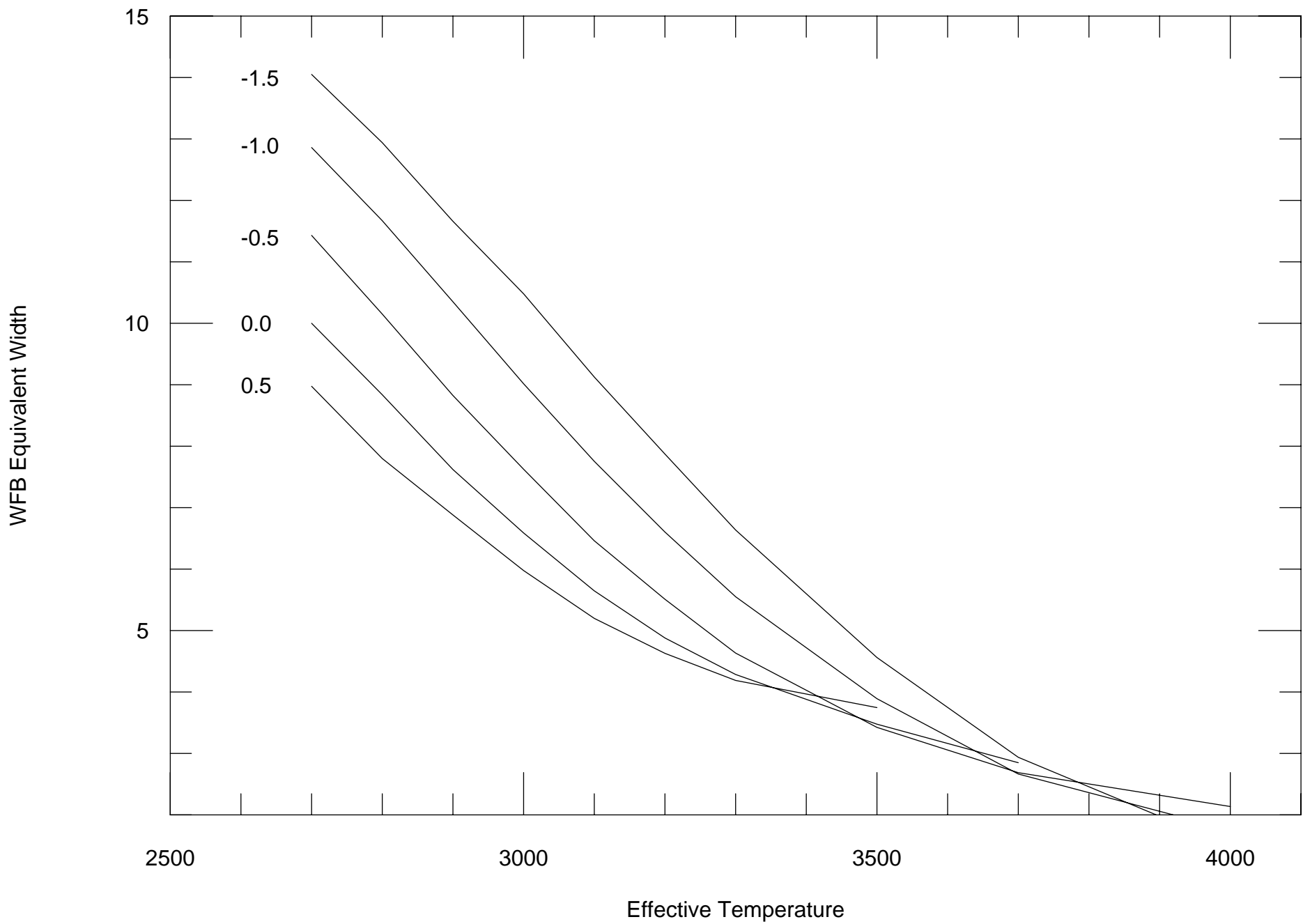
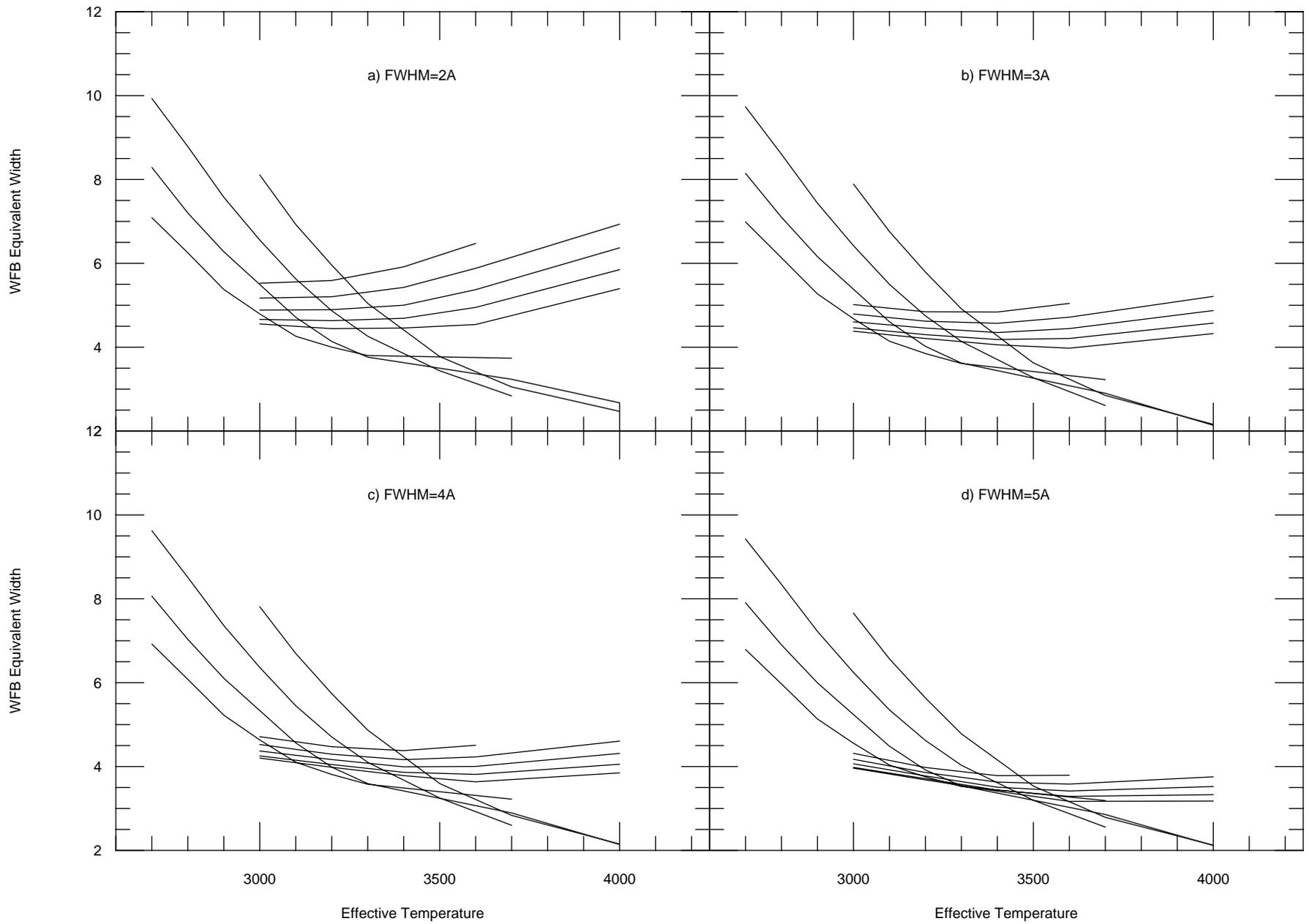
\end